\documentclass[aps,preprint]{revtex4}%
\usepackage{amsfonts}
\usepackage{amsmath}
\usepackage{amssymb}
\usepackage{graphicx}%
\begin{document}
\title{Enhancement of the Josephson current by magnetic field in superconducting
tunnel structures with paramagnetic spacer}
\author{V. N. Krivoruchko and E. A. Koshina}
\affiliation{Donetsk Physics \& Technology Institute NASU, R.Luxemburg Str., 72,
Donetsk-114, 83114 Ukraine}

\begin{abstract}
The dc Josephson critical current of a (S/M)IS tunnel structure in a parallel
magnetic field has been investigated (here S is a superconductor, S/M is the
proximity coupled S and paramagnet M bilayer and I is an insulating barrier).
We consider the case when, due to the Hund's rule, in the M metal the
effective molecular interaction aligns spins of the conducting electrons
antiparallel to localized spins of magnetic ions. It is predicted that for
tunnel structures under consideration there are the conditions when the
destructive action of the internal and the applied magnetic fields on Cooper
pairs is weakened and the increase of the\ applied magnetic field causes the
field-induced enhancement of the\ tunnel critical current. The experimental
realization of this interesting effect of the interplay between
superconductivity and magnetism is also discussed.


\end{abstract}
\maketitle

\section{Introduction}

In ferromagnetic\ (F) metals the exchange field $H_{E}$ , acting on the spin
of conducting electrons via the exchange interaction with magnetic moments of
ions, is in general so large as to inhibit superconductivity. When an external
magnetic field is applied, superconductivity is suppressed due to orbital and
spin pair breaking effects, as well. However, there are magnetic metals, such
as (EuSn)Mo$_{6}$S$_{8}$ [1,2] or\ HoMo$_{6}$S$_{8}$ [3], where the applied
magnetic field can induce superconductivity. Several mechanisms that may
enable superconductivity to develop in a ferromagnet or a paramagnet have been
investigated in more or less detail (see [4,5] and references therein). One of
them is the so-called Jaccarino-Peter effect [6]. It takes place in those
para- and ferro-magnetic metals, in which, due to Hund coupling energy, the
exchange interaction, $J\mathbf{sS}$ , orients the spins $\mathbf{s}$ of the
conducting electrons antiparallel to the spins $\mathbf{S}$ of rare earth
magnetic ions. The effective filed acting on the spin of conduction electron
is $\mu_{B}H+g\mu_{B}J<S>$ with $J<0$\ ($\mu_{B}$ is Bohr magneton, $g$ is
g-factor). In such magnetic metals the exchange field $g\mu_{B}J<S>$ can be
reduced by the external magnetic field $\mu_{B}H$ , so that the destructive
action of both fields on the conducting electrons can be weakened or even
canceled. If, in addition, these metals posses an attractive electron-electron
interaction, as, for example, in pseudoternary compounds [5], it is possible
to induce bulk superconductivity by a magnetic field.

In this report, we consider the dc Josephson effect for a tunnel structure
where one electrode is the proximity coupled bilayer of a superconducting film
(S) and a paramagnet (M) metal , while the second electrode is an S layer. The
system is under the effect of weak external magnetic field, which by itself is
insufficient to destroy superconductivity. The dc critical current of such a
junction has been calculated using approximate microscopic treatment based
on\ Gor'kov equations. We discuss the case when in the M metal the localized
paramagnetic moments of the ions, oriented by magnetic field, exert the
effective interaction on spins of the conducting electrons $J\mathbf{sS}$ .
The latter, whether it arises from the usual exchange interaction or due to
configuration mixing, according to Hund rules, is the antiferromagnetic type,
i.e. $J<0.$ In particular,\ such M metal could be a layer of pseudoternary
compounds like (EuSn)Mo$_{6}$S$_{8}$ or\ HoMo$_{6}$S$_{8}$. (While
experimentally the Jaccarino-Peter phenomenon was observed [1-5] for
paramagnets, this mechanism is applicable both to ferromagnetic and
paramagnetic metals, and both type of the magnetic orders will be assumed
here.) We demonstrate that in the region where the destructive action of the
fields on both tunnel electrodes is decreased, an increase of the\ magnetic
field causes the enhancement of the Josephson critical current.

\section{The model}

The system we are interested in is the (S/M)IS layered structure of the
superconducting S/M bilayer and S films separated by very thin insulating (I)
barrier (see Fig.1). The S/M bilayer consists of the proximity coupled
superconducting and paramagnet metals in good electric contact. It is assumed
that the thicknesses of the S layers are smaller than the superconducting
coherent length and that the thickness of the magnetic layer\ is smaller than
the condensate penetration length, i.e., $d_{S}<<\xi_{S}$\textbf{\ }and
$d_{M}<<\xi_{M}$. Here $\xi_{S(M)}$ is the superconducting coherence length of
the S(M) layer; $d_{S(M)}$\ is the thickness of\ the S(M) layer. In this case,
the superconducting order parameter may be regarded as being independent of
the coordinates and the influence of the magnetic layer on superconductivity
is not local. Other physical quantities characterizing the S/M bilayer are
modified, as well. Such an approach was recently discussed in\ [7,8] for SFIFS
structures, and, as was demonstrated, under these assumptions, a thin S/F
bilayer is equivalent to a superconducting ferromagnetic film with homogeneous
superconducting order parameter and an effective exchange field. Similarly, we
can consider the S/M bilayer as a thin SM film which is characterized by the
effective values of the superconducting order parameter $\Delta_{ef}$, the
coupling constant $\gamma_{ef}$ and the exchange field $H_{Eef}$ that are
determined by the following relations:%
\begin{align}
\Delta_{ef}/\Delta &  =\gamma_{ef}/\gamma=\nu_{S}d_{S}(\nu_{S}d_{S}+\nu
_{M}d_{M})^{-1}\text{ , }\\
H_{Eef}/H_{E}  &  =\nu_{M}d_{M}(\nu_{S}d_{S}+\nu_{M}d_{M})^{-1},
\end{align}
where $\nu_{S}$ and $\nu_{M}$ are the densities of quasiparticles states in
the superconductor and magnetic metals, respectively; $\gamma$\ is the
coupling constant in the S metal. We emphasize that \textit{the
superconductivity of the M metal is due to proximity effect}. The applied
magnetic field is too weak to induce the superconducting properties through
the Jaccarino-Peter scenario, if the M metal is the pseudoternary compound.
While in the latter case the M metal can posses a nonzero electron-electron
interaction, we will neglect this interaction assuming for the M layer a
vanishing value of the bare superconducting order parameter $\Delta_{M}^{0}=0$
, so that relation (1) still remains valid.

The system is under the effect of parallel magnetic field $H$. We will also
assume that the thicknesses of the SM and S films are smaller than the London
penetration depth $\lambda_{SM}$ and $\lambda_{S}$ , correspondingly. Then the
magnetic field is homogeneous in both electrodes. The conditions $d_{S}%
<<\xi_{S}$\textbf{\ , }$d_{M}<<\xi_{M}$ ensure that the orbital effects can be
neglected, as well. The longitudinal dimension of the junction, $W$, is
supposed to be much less than the\ Josephson penetration depth, $W<<$
$\lambda_{J}$ , so that a flux quantum can not be trapped by the junction:
$HW(d_{M}+2d_{S}+t)<<\phi_{O}$\ , here $\phi_{O}$ is the flux quantum, $t$ is
the thickness of the insulator.

If the transparency of the insulating layer is small enough, we can neglect
the effect of a tunnel current on the superconducting state of the electrodes
and use the relation of the standard tunnel theory [9], according to which the
distribution of the Josephson current density $j_{T}(x)$ flowing in the
z-direction through the barrier (see Fig.1) takes the form $\ j_{T}%
(x)=I_{C}\sin\varphi(x)$. Here $\varphi(x)$ is the phase difference of the
order parameter across the barrier,\ while the Josephson current density
maximum $I_{C}$ is determined by the properties of the electrodes. In this
report we present the results of the calculation of the critical current
$I_{C}$ for the tunnel junction under consideration.

\section{Critical current}

As far as the exchange field and the external magnetic field act only on\ the
spin of electrons we can write the Gor'kov equations for the S and SM layers
in the magnetic field in the form:\
\begin{align}
(i\varepsilon_{n}+\xi-\sigma H_{S(SM)}\mathbf{)}\hat{G}_{\varepsilon
S(SM)}+\hat{\Delta}_{S(SM)}\hat{F}_{\varepsilon S(SM)}^{+}  &  =1,\\
(-i\varepsilon_{n}+\xi-\sigma H_{S(SM)}\mathbf{)}\hat{F}_{\varepsilon
S(SM)}+\hat{\Delta}_{S(SM)}\hat{G}_{\varepsilon S(SM)}  &  =0,
\end{align}
where\ $\xi=\varepsilon(p)-\varepsilon_{F}$, \ $\varepsilon_{F}$ is the Fermi
energy, $\varepsilon(p)$ is the quasiparticle spectrum, $\sigma=\pm1$\ ,
$\varepsilon_{n}=\pi T(2n+1)$, $n=0,\pm1,\pm2,\pm3,...$ are Matsubara
frequencies; $T$ is the temperature of the junction (here and below we have
taken the system of units with $\hbar=\mu_{B}=k_{B}=1$); $H_{SM}=H_{Eef}-H$ is
the resulting magnetic field in the SM bilayer (the subscript $SM$) and
$H_{S}=$\ $H$ is the magnetic field in the S layer (the subscript $S$) ;
$G_{\varepsilon}$ and $F_{\varepsilon}$ are normal and anomalous Green
functions. The equations are also supplemented with the well known
self-consistency equations for the order parameters. In the case of
conventional singlet superconducting pairing, when $\hat{\Delta}=i\sigma
_{y}\Delta$ ($\sigma_{y}$\ is Pauli matrix), one can easily find (see, e.g., [8]):%

\begin{equation}
\ln\left(  \frac{\Delta_{0}}{\Delta_{S(SM)}}\right)  =%
%TCIMACRO{\tint _{0}^{\omega_{D}}}%
%BeginExpansion
{\textstyle\int_{0}^{\omega_{D}}}
%EndExpansion
\frac{dx}{\sqrt{x^{2}+\Delta_{S(SM)}^{2}}}\{\frac{1}{\exp[\beta\sqrt
{x^{2}+\Delta_{S(SM)}^{2}}-H_{S(SM)}]+1}+
\end{equation}%
\[
+\frac{1}{\exp[\beta\sqrt{x^{2}+\Delta_{S(SM)}^{2}}+H_{S(SM)}]+1}\}
\]
where $\Delta_{0}=\Delta(0,0)$ is the BCS gap at zero temperature and in the
absence of both the applied and the exchange fields; $\omega_{D}$ is the Debye
frequency; $\beta=1/T$ ; $\Delta_{SM}(T,H_{SM}),$ $\Delta_{S}(T,H_{S})$ are
the superconducting order parameters of the SM and S electrodes, respectively.
If $H_{S(SM)}=0$, formula (5) is reduced to Eq. (16.27) of Ref. 10.

In accordance with the Green's function formalism, the critical current of the
SMIS junction can be written as follows:%
\begin{equation}
I_{C}=(2\pi T/eR_{N})Sp\sum_{n,\sigma}f_{SM}(H_{SM})f_{S}(H_{S}),
\end{equation}
where $R_{N}$ is the contact resistance in the normal state and
$f_{\varepsilon SM(S)}$ are averaged over energy $\xi$ anomalous Green
functions. From Eqs. (3) and (4) one can easily find that:%
\begin{equation}
f_{\varepsilon SM(S)}=\Delta\lbrack(\varepsilon_{n}+i\sigma H_{SM(S)}%
)^{2}+\Delta^{2}]^{-1/2}.
\end{equation}
Using Eqs. (6) and (7), after summation over spin index, we find for the
reduced (i.e. $eR_{N}\{4\pi T\Delta_{0}^{2}\}^{-1}I_{C}$ ) quantity
\begin{align}
j_{C}(T,H)  &  =\Delta_{SM}(T,H_{SM})\Delta_{S}(T,H)\Delta_{0}^{-2}%
\times\nonumber\\
&  \operatorname{Re}\sum_{n}\{[(\varepsilon_{n}-i(H_{Eef}-H))^{2}+\Delta
_{SM}^{2}(T,|H_{Eef}-H|)][(\varepsilon_{n}+iH)^{2}+\Delta_{S}^{2}%
(T,H)]\}^{-1/2}%
\end{align}
The Josephson critical current of the junction, as function of the fields and
temperature, can be calculated using formula (8) and self-consistency equation
(5). In the general case, the dependence of the superconducting order
parameter on effective field can be complex enough due to the possibility of
transition to the nonhomogeneous (Larkin-Ovchinnikov-Fulde-Ferrell) phase
[11,12]. We will not touch upon this scenario here, restricting the
consideration below to the region with the homogeneous superconducting state.
Even in this case at arbitrary temperatures the values of the $\Delta
_{SM}(T,|H_{Eef}-H|)$ and $\Delta_{S}(T,H)$ can be determined only
numerically. The phase diagram of a homogeneous superconducting state in the
$H-T$ plane has been obtained earlier\ (see, e.g., [8]). At finite
temperatures, it is found that $\Delta(T,H)$ has a sudden drop from a finite
value to zero at a threshold of $H$, exhibiting a first-order phase transition
from a superconducting state to a normal state. Using these results, from Eq.
(5) we take only one branch of solutions, corresponding to a stable
homogeneous superconducting state.\ It should be also noted that, as far as
$H_{E}\varpropto<S>$ , a self-consistency equation should be used for
$H_{Eef}$, as well. However, we will suppose that $H_{Eef}$ , being much
smaller than in isolated M film, is still larger than $\Delta_{SM}%
(T,|(H_{Eef}-H\mathbf{)}|)$ for full temperature region of the homogeneous
superconducting state. So, proceeding in the way to tackle the new physics, we
will ignore the temperature dependence of the $H_{Eef}$ in Eq. (8).

Figures 2 and 3 show the results of numerical calculations of expression (8)
for the Josephson critical current versus external magnetic field for the case
of low $T=0.1T_{C}$\ and finite $T=0.7T_{C}$\ \ temperatures, and different
values of the exchange field. To keep the discussion simple, for the SM and S
layers we put $\Delta_{SM}(0,0)=\Delta_{S}(0,0)=\Delta_{0}$ . As is seen in
the figures, for some interval of the applied magnetic field the enhancement
of the dc Josephson current takes place in comparison with the case of
$\ H=0$. Note that, the larger the effective field $H_{Eef}$ is, the larger
growth of the critical current can be observed (compare, for example, the
$j_{C}\ $curves for $H_{Eef}$ = $0.4\Delta_{0}$ and $H_{Eef}$ = $0.6\Delta
_{0}$\ at $H=0$ in Fig. 2). This behavior is also predicted by expression (8).
A sudden break off in the $j_{C}(H)\ $dependences in the presence of $H$
results due to a first-order phase transition from a superconducting state
with finite $\Delta(T,H)$ to a normal state with $\Delta(T,H)=0$ .

\section{Discussion}

As is well known [13,14], due to the difference in energy between spin-up and
spin-down electrons and holes under the exchange field of a ferromagnet, a
singlet Cooper pair, adiabatically injected from a superconductor into a
ferromagnet, acquires a finite momentum. As a result, proximity induced
superconductivity of the F layer is spatially inhomogeneous and the order
parameter contains nodes where the phase changes by $\pi$. Particularly,
transport properties of tunnel SF structures have turned out to be quite
unusual. The $\pi$ state is characterized by the phase shift of $\pi$ in the
ground state of the junction and is formally described by the negative
critical current I$_{C}$ in the Josephson current-phase relation:
$j(\varphi)=I_{C}\sin(\varphi)$ . The\ $\pi$-phase\ state of an SFS weak link
due to Cooper pair spatial oscillation was first predicted by Buzdin
\textit{et al}., [15,16].\ Experiments that have been performed by now on SFS
weak links [17,18] and SIFS\ tunnel junctions [19]\ directly prove the $\pi
$-phase superconductivity.

There is another interesting case of a thin F layer, $d_{F}<<\xi_{F}$, being
in contact with an S layer. As far as the thickness of the F layer $d_{F}$ is
much less then the corresponding superconducting coherence length $\xi_{F}$
there is spin splitting but there is no order parameter oscillation in the F
layer. Surprisingly, but it was recently predicted [7,8,20-24] that for SFIFS
tunnel structures with very thin F layers one can, on condition of parallel
orientation of the F layers magnetization, turn the junction into the $\pi
$-phase state with the critical current inversion; if the F layers internal
fields have antiparallel orientation, one can even enhance the tunnel
current.\ It is obvious, that physics behind the inversion and the enhancement
of the supercurrent in this case differs from that proposed by Buzdin
\textit{et al.}\ Namely, in this case the $\pi$-phase state is due to
superconducting phase jump at the SF interface [21,24 ]. The exchange-field
enhancement of the critical current for SFIFS tunnel structure can be
qualitatively understood using the simple fact that the Cooper pairs consist
of two electrons with opposite spin directions. Pair--breaking effects due to
spin-polarized electrons are weaker in the antiparallel-aligned configuration
since spin polarizations from the exchange fields of the F layers are of
opposite signs and at some conditions can cancel each other. More formally,
one can show that the maximum of the supercurrent is achieved exactly at those
values of the exchange field when two singularities in the quasiparticle
density of states overlap [23].

We emphasize that the scenario of the magnetic-field enhancement of the
critical current discussed here, differs from those studied before for SFIFS
tunnel structures. In our case the pair--breaking effect due to spin-polarized
electrons is weakened in\ the SM electrode since the spin polarizations from
the exchange field of the magnetic ions and the applied field are of opposite
signs and reduce each other. On the other hand, the paramagnetic effect
induced by the external field is increased for the Cooper pairs of the S
electrode if the applied field is increased. Competition of these two opposite
effects determines the critical current behavior for the SMIS junction in the
magnetic field. In our case the mechanism described above is valid for full
temperature region of the homogeneous superconducting state (see, e.g., Fig.
3), while for the SFIFS system with antiparallel geometry - only at low
temperature $T<<T_{C}$\ [7,8].

In conclusion, we calculate the dc critical current of the (S/M)IS tunnel
structure, where one electrode is the proximity coupled bilayer of a
superconducting film and a paramagnet metal, while the second electrode is an
S layer. The structure is under the effect of weak parallel external magnetic
field. In the magnetic metal the localized magnetic moments of the ions,
oriented by the magnetic field, exert the effective interaction on spins of
the conduction electrons $J\mathbf{sS}$. The latter, whether it arises from
the usual exchange interaction or due to configuration mixing, according to
the Hund rules, is the antiferromagnetic type, i.e. $J<0$ . In
particular,\ such a film can be the layer of the pseudoternary compounds like
(EuSn)Mo$_{6}$S$_{8}$,\ HoMo$_{6}$S$_{8}$, etc. There are no specific
requirements on the superconductor, so that it can be any superconducting film
proximity coupled with the magnetic metal. Using approximate microscopic
treatment of the S/M bilayer and the S layer, we have predicted the effect of
magnetic-field-induced supercurrent enhancement in the tunnel structure. This
striking behavior contrasts with the suppression of the critical current by
magnetic field. The idea to use a magnetic material in which the effective
magnetic interaction aligns spins of the conducting electrons antiparallel to
the localized spin of magnetic ions, in order to enhance superconductivity of
superconductor-magnetic metal multilayered structures, has not been considered
before and, to our best knowledge, is new. The existing large variety of
magnetic materials, the ternary compounds in particular, should allow
experimental realization of this interesting new effect of the interplay
between superconducting and magnetic orders.

We thank Dr. M. Belogolovskii for useful discussions.

\begin{center}
\newpage

Figure captions
\end{center}

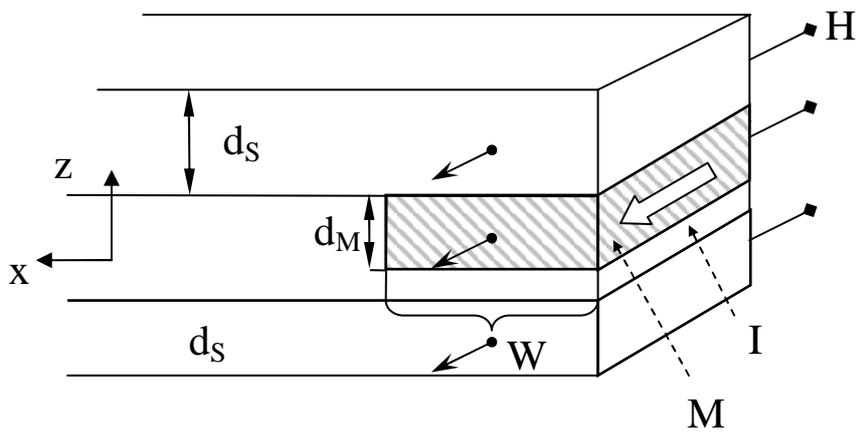
FIG. 1. (S/M)IS system in a parallel magnetic field. Here S is a
superconductor; M is a magnetic metal; I is an insulating barrier; W is
longitudinal dimension of the junction.

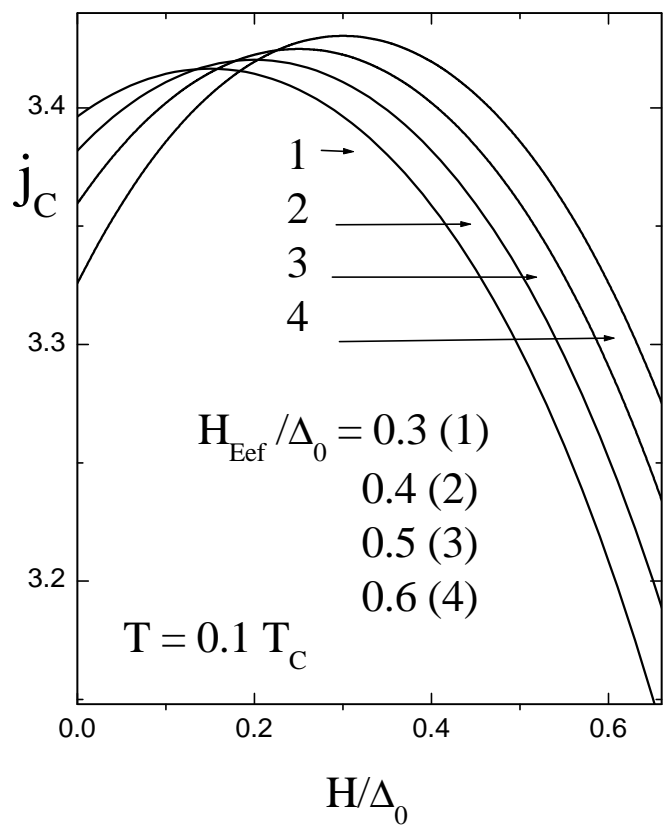
FIG. 2.\ Critical current of the SMIS tunnel junction vs external magnetic
field for $T=0.1T_{C},$ $\Delta_{SM}(0,0)=\Delta_{S}(0,0)=\Delta_{0}$ and
different values of the effective exchange field in the SM bilayer:
$H_{Eef}/\Delta_{0}=0.3,$ $0.4,$ $0.5$ and $0.6$ (curves 1, 2, 3 and 4, respectively).

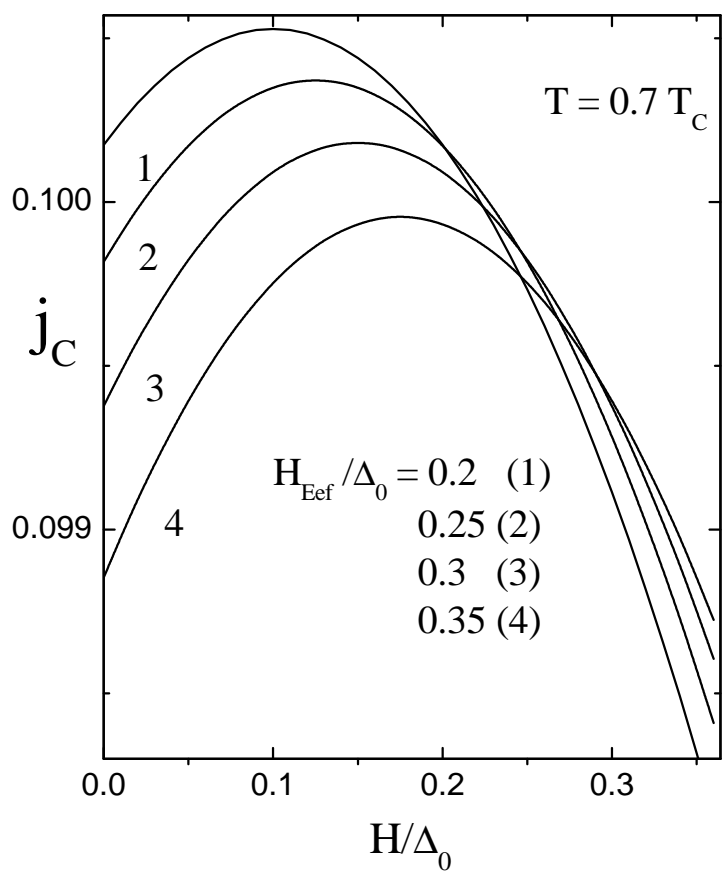
FIG. 3. Critical current of the SMIS tunnel junction vs external magnetic
field for $T=0.7T_{C},$ $\Delta_{SM}(0,0)=\Delta_{S}(0,0)=\Delta_{0}$ and
different values of the effective exchange field in the SM bilayer:
$H_{Eef}/\Delta_{0}=0.2,$ $0.25,$ $0.3$ and $0.35$ (curves 1, 2, 3 and 4, respectively).

\end{document}